\documentclass[pra,bibnotes,twocolumn]{revtex4}
\usepackage[dvips]{pstcol}
\usepackage{pst-node}
\usepackage{pst-plot}
\usepackage{pst-coil}
\usepackage{graphicx}
\begin{document}
\draft
\def\ds{\displaystyle}
\title{Self-accelerating Parabolic Cylinder Waves in 1-D}
\author{ C. Yuce}
\address{Department of Physics, Anadolu University, Turkey }
\email{cyuce@anadolu.edu.tr}
\date{\today}
\begin{abstract}
We introduce a new self-accelerating wave packet solution of the Schrodinger equation in one dimension. We obtain an exact analytical parabolic cylinder wave for the inverted harmonic potential. We show that truncated parabolic cylinder waves exhibits their accelerating feature.   
\end{abstract}
\maketitle

\section{Introduction}

Airy wave, which is a stationary solution of the Schrodinger equation in the presence of a linear potential was surprisingly shown to be an exact nondiffracting wave for the free particle Schrodinger equation \cite{berry}. In addition to nondiffracting character, the Airy wave was also shown to be self-accelerating, which implies that the Airy wave breaks the Ehrenfest theorem. This is because of the non-integrability of the Airy function. Long after this theoretical prediction by Berry and Balasz, an experimental observation of truncated Airy wave packet was achieved within
the context of optics, where the paraxial equation of diffraction in optics and the Schrodinger equation are equivalent \cite{deney1,deney2,deney3,deney4,deney5,ek}. In a couple of years after its first experimental realization in optics, the accelerating Airy wave was also obtained for free electrons using a nanoscale hologram
technique \cite{electron}. Note that although the Airy wave is truncated, which is necessary for its experimental realization, its non-diffracting and accelerating feature are not lost within a large distance. In fact, the main lobe of the truncated Airy waves was found to follow a parabolic trajectory. The experimental realization of Airy waves triggered other experimental and theoretical investigations on other types of accelerating beams. In addition to Airy waves, Weber and Mathieu beams \cite{matweb1,matweb2,matweb3} and accelerating beams in curved space \cite{curvedMax} were found. Self accelerating waves for nonlinear paraxial \cite{nonlin1,nonlin2,nonlin3,nonlin4} and nonparaxial equation \cite{nonparax1,nonparax2} were explored. Self-accelerating waves were investigated by some authors in nonuniform systems such as matter waves in an expulsive potential \cite{cemmodern}, Airy beams in a parabolic potential \cite{ yigi}, nondiffracting Bessel waves in optical antiguides \cite{antik} and a periodical photonics system \cite{onemli01,onemli02,onemli03,00102,102,9,10,101}. In the latter system, accelerating Wannier-Stark states in optical lattices has been theoretically proposed and experimentally realized \cite{onemli01,onemli02}. It was observed that the two main lobes of Wannier-Stark states experience self-bending along opposite trajectories. It was shown that diffractionless and accelerating beams can only exist in a propagation distance dependent periodic potentials \cite{onemli03}. In a separate work, propagation of waves that has Airy intensity profile was investigated in nonuniform media \cite{101}. We note that different types of caustic trajectories can be generated in a periodical lattice by appropriately engineering the initial phase \cite{9}. \\
In 1-D, Airy wave is the only exactly known non-diffracting self-accelerating wave in quantum physics. A question arises. Can we find other exact self-accelerating waves for systems described by the 1-D Schrodinger equation? The topic of self accelerating waves in the presence of potential has not been understood yet. To obtain an ideal self-accelerating wave, we must analytically solve the Schrodinger equation. In this paper, we find another exact analytical self-accelerating solution of the 1-D Schrodinger equation. The potential we consider is time-dependent inverted harmonic potential and the form of the new self-accelerating is of the parabolic cylinder function. As a special case, we derived diffracting self-accelerating wave for a free particle and show that the range at which the wave remains shape-preserving for truncated parabolic cylinder wave is as large as that of the truncated Airy wave. We also discuss self-accelerating solution in momentum space. 

\section{Parabolic Cylinder Waves}

We start with the 1-D Schrodinger equation  for the inverted harmonic potential. (i.e., the harmonic force is repulsive rather than attractive \cite{ceminverted,inverted2,onofrio,ekkk01,ekkk02,ekkk03}). The equation reads 
\begin{equation}\label{cn10}
-\frac{1}{2}\frac{\partial^2
\psi}{\partial x^2}-\frac{\omega^2}{2 }x^2~\psi=i\frac{{\partial}\psi}{\partial  t}
\end{equation}
where $\omega^2(t)$ is a time dependent angular frequency and $m=\hbar=1$. Note that the paraxial equation in optics is obtained if we replace time with the propagation distance.\\
According to the Ehrenfest's theorem, the acceleration of a wave packet in our system can be
found using the equation $\ds{\frac{d^2<x>}{dt^2}-\omega^2 <x>=0}$. The Ehrenfest's theorem shows that the expectation value obeys the classical dynamical laws. Therefore the acceleration of the wave packet matches the acceleration of a classical particle moving in one dimension under the influence of the inverted harmonic potential. This statement is true as long as the wave packet is square integrable. In the case of non-integrable wave packets, the Ehrenfest's theorem can not be applied and hence the above statement for the inverted oscillator potential may not hold. In fact, there is no general theory for nonintegrable wave packets. We emphasize that two different nonintegrable wave packets in a given potential can have different self-accelerations. We should solve the Schrodinger equation analytically to get a formula for self-acceleration since numerical computation can only be performed for truncated waves, which satisfy the Ehrenfest's theorem. It is the purpose of this paper to find the self-accelerating solution for the inverted harmonic potential.\\ 
Let $D_n(x)$ be the parabolic cylinder function and $\omega_0>0$ be an arbitrary real constant. The two linearly independent self-accelerating solutions of the Schrodinger equation (\ref{cn10}) are given by (see the Methods section)
\begin{eqnarray}\label{ufegdsdaw2b}
\psi_{n}=\frac{1}{\sqrt{L}}\exp{\left( i(\dot{x}_c {(x-x_c)}+
\frac{\dot{L}}{2L}
{(x-x_c)}^2+S-\int\frac{E}{L^2}dt)     \right)}\nonumber\\
~ D_{  \lambda_n    }  \left( \sqrt{2}(-1)^{\frac{2n-1}{4}}    ( \frac{ a_0}{\omega_0^{3/2}}+\sqrt{\omega_0}~ \frac{x-x_c}{L}     ) \right) 
\end{eqnarray}
where $\ds{n=1,2}$ and $\ds{\lambda_1=  \frac{{}i a_0^2-2i E \omega_0^2-\omega_0^3}{2\omega^3}     }$ and $\ds{\lambda_2=  \frac{{-}i a_0^2+2i E\omega_0^2-\omega_0^3}{2\omega_0^3}     }$, $\omega_0$ and $\ds{a_0}$ are arbitrary real valued
constants, $\ds{\dot{S}(t)=\frac{1}{2} \dot{x_c}^2-\frac{\omega^2}{2}
x_c^2}$ and $E$ is the energy in the limit $\omega^2=\omega_0^2$, $x_c=0$ and $L=1$. We stress that there are two free parameters, $a_0$ and $\omega_0$ in our solution. The constant $\omega_0$ determines how the width of the wave packet changes in time and the constant $a_0$ plays a vital role on the self acceleration as will be shown below. In our solution, the time-dependent function
$\ds{x_c(t)}$ describes translation of the wave packet with the acceleration $\ds{\ddot{x_c}}$ and $L(t)$ is a time-dependent dimensionless scale
factor, where we initially take
$L(0)=1$. The equations satisfied by these two functions are derived in the Methods section. They are given by
\begin{eqnarray}\label{udenklemiyeni}
\ddot{L}-\omega^2 ~L&=&-\frac{\omega_0^2}{L^3},\\\label{udenklemidjlaadx}
\ddot{x}_c-\omega^2~x_c&=&-\frac{a_0}{L^3},
\end{eqnarray}
As can be easily seen from (\ref{ufegdsdaw2b}), our solutions are stationary in the moving coordinate system with $\ds{x^{\prime }=\frac{x-x_c}{L}}$ while self-accelerating in the original coordinate system. The self-accelerating feature can also be seen from the right hand side of the equation (\ref{udenklemidjlaadx}). The term with $a_0$, which is absent in the Ehrenfest's theorem is responsible for the self-acceleration. This inconsistency is due to the fact that the exact solutions (\ref{ufegdsdaw2b}) are not square integrable. \\
Notice the difference between our solutions and the one obtained by Berry and Balasz, which has only one self-accelerating solution \cite{berry}. A combination of the two linearly independent solutions (\ref{ufegdsdaw2b}) is again a self-accelerating solution. If the two superposing waves have the same $a_0$ and $\omega_0$, then the superposed wave moves with the same self-acceleration. If they have different $a_0$ and $\omega_0$, then a complex motion occurs. In fig-1, we plot intensities for $\psi_1(x)$ (a), $\psi_2(x)$ (b) when $E=0$. For large values of $E$, the intensities of them are very close to each other (c). As can be seen, $\psi_1$ and $\psi_2$ don't converge in the negative and positive branches, respectively, while a combination of them has oscillating tails in both directions. We emphasize that the term $a_0$ is practically determined by the position of the main lobe of the parabolic cylinder function (\ref{ufegdsdaw2b}). Therefore acceleration of the parabolic cylinder waves can be changed by varying the initial position of the main lobe of the parabolic cylinder function. This result is interesting in the sense that acceleration occurs not due to the external force but due to the initial form of the wave function. This is the reason why such waves are called self-accelerating waves.\\
Having obtained the exact self-accelerating solutions, let us now discuss the equations for $L(t)$ and $x_c(t)$ in detail. The equation for $L$ implies that
spreading of this wave-packet is determined
by the two angular frequencies $\omega$ and $\omega_0$. The latter equation implies that the acceleration of the wave packet depends on the external inverted harmonic potential and the free parameter $a_0$, which is fixed by the position of the initial main lobe. As a special case, consider the free particle with $\ds{\omega=\omega_0=0}$. In this case, the wave packet is diffraction-free since the width of the wave packet remains constant, i.e., $L(t)=1$ and the corresponding self acceleration is constant, $\ds{\ddot{x_c}=-a_0}$. So we say that self-acceleration formula of Airy wave \cite{berry} can be recovered in this special case. Another self-accelerating wave packet for the free particle can be obtained if we choose $\ds{\omega=0}$ but $\ds{\omega_0\neq0}$. In this case self-acceleration occurs in free space while the width of the wave packet changes. The solution of the equation (\ref{udenklemiyeni}) is given by $\ds{L(t)=\sqrt{1+2t\sqrt{\omega_0^2+\omega_1^2}+\omega_1^2t^2}}$, where $\omega_1$ is an arbitrary constant. The acceleration of the wave packet, $\ds{\ddot{x}_{c}}$, is given by $\ds{-\frac{a_0}{{L(t)}^{3} }    }    $, which indicates that the self-acceleration is not constant in time and practically goes to zero at large times. We conclude that  unlike the nondiffracting self-accelerating Airy wave, the parabolic cylinder wave diffracts while self-accelerating in free space. We say that the two different nonintegrable waves, Airy and parabolic cylinder waves, break the Ehrenfest's theorem and they have different self-acceleration. We stress that quasi-diffractionless self-accelertating wave packet can be practically realized if we choose a small value of $\omega_0$ and $\omega_1=0$. In practice, not the infinite-energy self-accelerating but a truncated wave packet can be realizable. We can perform numerical computation for a truncated parabolic cylinder wave. Suppose the following initial truncated wave packet 
\begin{equation}\label{9eiwfju}
\Psi_{i}=N\exp{(-\epsilon x^2)}\psi_i
\end{equation}
where $N$ is the normalization factor, $\ds{\epsilon<<1}$ is a small parameter that makes the wave packet square integrable and $\psi_i$ are defined in (\ref{ufegdsdaw2b}). Note that only the ideal wave with an infinite energy violates the Ehrenfest's theorem and self-acceleration occurs no matter how far such a wave packet moves. Fortunately, the truncated wave still exhibits accelerating behavior within a limited propagation distance. In our numerical computation, we start with the initial wave $\ds{\Psi_1(x,0)-\Psi_2(x,0)}$ with the parameters $\ds{\omega_0=a_0=1}$ and the decay parameter $\ds{\epsilon=1/20}$. The Fig.2 (a) shows the density plot, $\ds{|\Psi_1(x,t)-\Psi_2(x,t)|^2}$. As can be seen from the figure, two main lobes accelerate towards each other and then move opposite to each other. We note that a similar effect was observed in a discrete system. The two main lobes of the Wannier-Stark beams was experimentally demonstrated to be self-bended along two opposite trajectories in a uniform one-dimensional photonic lattice \cite{onemli02}. Our system is different from \cite{onemli02} since it is continuous and diffraction occurs. As a second example, we want quasi-diffractionless propagation in the free space. So, we consider $\ds{a_0=5\omega_0=1}$ and start with $\Psi_2$ with $\ds{\epsilon=\frac{1}{100}}$ and $\dot{L}(0)=0$. The Fig.2 (b) shows the density plot. As can be seen, the truncated parabolic cylinder wave still exhibits accelerating behavior. Compare it to the propagation of the well-known truncated Airy wave $\ds{e^{0.1x} Ai(2^{1/3}x)}$ in the Fig.2 (c). We conclude that the distance that accelerating behavior for the truncated parabolic wave can be observed before diffraction takes over is almost as large as the distance for the truncated Airy wave. \\
Self-accelerating solution for time-dependent inverted oscillator can be found for a given $\ds{\omega^2(t)}$ by solving (\ref{udenklemiyeni}) and (\ref{udenklemidjlaadx}) together. Here we consider a couple of special cases. Consider first the inverted harmonic potential with a constant angular frequency, $\omega^2=\omega_0^2$. In this case, one particular solution of (\ref{udenklemiyeni}) is given by $L(t)=\sqrt{\sqrt{1-4\omega_0^2c^2}+2 c \omega_0\sinh{(2\omega_0t)}}$, where $\ds{c<1/4\omega_0^2}$ is an arbitrary constant. The equation ({\ref{udenklemidjlaadx}) admits the solution $\ds{x_c(t)= \frac{a_0}{\omega_0^2}~L(t) +x_0 \sinh{\omega_0 t} }$, where the first term is the self-acceleration and the second term is due to the external inverted harmonic potential. If $c=0$, i.e., $L(0)=1$, then nondiffracting wave packet solution is obtained since the width of the wave packet is constant at all time. In this case, the corresponding wave packet is not self-accelerating since the Ehrenfest's theorem gives the same result although the wave packet is non-integrable. If $c\neq0$, self-accelerating and diffracting wave packet can be obtained for time-independent inverted harmonic potential. As discussed above, one has to start with a truncated wave (\ref{9eiwfju}) for experimental purposes. In this way, we can observe the effect of self-acceleration within a limited distance. We note that the inverted harmonic potential is maximum at the origin and has no minimum points (it goes to minus infinity as $\ds{x\rightarrow\mp  \infty}$). So, we expect that a wave packet moves in the positive direction indefinitely if the initial expectation value $\ds{<x(t=0)>}$ is positive and vice versa. The self-acceleration may change this picture and makes it possible for the wave packet to move towards the origin, where the inverted harmonic potential is maximum. We can design our system to see this effect for the truncated wave. Let us choose $a_0=-5$ and shift the center of the main lobe away from the center of the inverted harmonic potential ($\ds{<x(t=0)>}$ is positive). The Fig.3 (a)-(c) show the density plot for our truncated wave and a Gaussian wave. They are both shifted away from the center of the inverted harmonic potential. The Gaussian beam accelerates in the positive direction as it is expected. However, our truncated wave in Fig.1 (b) exhibits its accelerating feature and moves towards the point where the potential is maximum. This is because the self-acceleration overcomes the acceleration due to the inverted harmonic potential. If the two accelerations cancel each other, the main peak of the accelerating beam becomes motionless as if there exist no external potential. \\
Consider finally another interesting case with $\ds{a_0=\omega_0^2}$ but $\omega^2(t)$ is a time dependent function. In this case, there exist a solution where the scale function and $\ds{x_c}$ have the same form, i.e., $\ds{L(t)=x_c(t)}$ as can be seen from (\ref{udenklemiyeni},\ref{udenklemidjlaadx}). As an example of this case, consider a self-oscillating and breathing wave packet with $\ds{L=x_c=1+\epsilon \sin{{\omega_0}t}}$, where $\ds{\epsilon}$ is a small parameter. To obtain such a solution, we need $\ds{\omega^2(t)=\frac{\omega_0^2}{(1+\epsilon \sin{{\omega_0}t})^4}- \frac{\epsilon \Omega^2\sin{\omega_0   t}}{1+\epsilon \sin{\omega_0  t}}   }$. \\
To this end, let us mention self-accelerating waves in momentum space. The Schrodinger equations in $x$-space and momentum space have the same form for harmonic potentials. Let $\phi(p)$ be the Fourier transform of the wave function $\psi(x)$. Then the Schrodinger equation in momentum space reads
$\ds{H\phi(p)=i\frac{{\partial}\phi}{\partial  \tau}}$, where $\ds{H=- \frac{1}{2 }  \frac{\partial^2}{\partial p^2}-\frac{1}{2\omega^2}  p^2}$ and $\ds{\tau=-{\omega}^2t  }$. The solution in momentum space can be obtained if we replace $\ds{x\rightarrow{p}}$, $\ds{x_c\rightarrow{p_c}}$, $\ds{\omega(t)\rightarrow\frac{1}{\omega(t)}}$ and $\ds{t\rightarrow-{\omega}^2t  }$ in the solution (\ref{ufegdsdaw2b}).\\
To sum up, we have introduced a new type of self-accelerating waves. Airy waves were known to be an exact self-accelerating solution in free space in 1-D. We have shown that parabolic cylinder waves are the exact self-accelerating solution for the 1-D inverted harmonic potential. We have shown that truncated parabolic cylinder waves  resists the effects of diffraction for long distances in free space.\\
This study was supported by Anadolu University Scientific Research Projects Commission under the grant no: 1605F344.

\section{Appendix}

Here, we derive two linearly independent self-accelerating solutions (\ref{ufegdsdaw2b}). To get self-accelerating wave packet solution, let us first make a coordinate transformation $\ds{x\rightarrow  \frac{x-x_c}{L}  }$ in (\ref{cn10}), where the time dependent function $x_c(t)$ describes translation and $L(t)$ is a time dependent dimensionless scale factor to be determined later. Under this coordinate transformation, the time derivative operator transforms as $\ds{\partial_t\rightarrow\partial_t-L^{-1} (\dot{L}  x^{\prime}+\dot{x_c } ) \partial_{x^{\prime}}}$, where dot denotes derivation with respect to time. In the accelerating frame, we will seek
the solution of the form 
\begin{equation}\label{s3fioskd}
\psi(x^{\prime},t)=\exp{\left(i(\alpha
{x^{\prime}}+ \frac{\beta}{2}{x^{\prime}}^2+S)\right)}~ \frac{\phi({x^{\prime}},t) }{\sqrt{L}}
\end{equation}
where $\alpha$, $\beta$ and $S$ are time-dependent functions to be determined. Substitute this ansatz into the corresponding Schrodinger equation and demand that the resulting equation includes time-dependent harmonic and linear potential terms. Therefore we choose $\ds{\alpha(t)=L\dot{x_c}}$ , $\ds{\beta(t)= L\dot{L}}$ and $\ds{\dot{S}=\frac{1}{2} \dot{x_c}^2-\frac{\omega^2}{2}
x_c^2}$. The resulting equation reads
\begin{equation}\label{revfckdsj}
-\frac{1}{2L^2}\frac{\partial^2\phi }{\partial {x^{\prime}}^2} +(\frac{1}{2}\Omega^2 {x^{\prime}}^2 +U {x^{\prime}})\phi =i\frac{\partial\phi }{\partial t}
\end{equation}
where $\ds{\Omega^2=L(\ddot{L}-\omega^2 L) }$ and $\ds{U=L(\ddot{x_c}-\omega^2 x_c)}$. Suppose $L$ and $x_c$ satisfy the equations (\ref{udenklemiyeni},\ref{udenklemidjlaadx}), respectively and $\ds{\phi({x^{\prime}},t)=e^{-i\int \frac{E}{L^2}dt}\phi({x^{\prime}}) }$, where $E$ is a constant. We further demand that the resulting equation includes time-independent inverted harmonic oscillator plus linear potential terms. So, we choose $\Omega^2=-\omega_0^2/L^2$ and $U=-a_0/L^2$, where $\omega_0>0$ and $a_0$ are arbitrary real constants.  As a result, we obtain  the equations (\ref{udenklemiyeni},\ref{udenklemidjlaadx}) and the transformed time-independent Schrodinger equation
\begin{equation}\label{revffuhdjz}
-\frac{1}{2 }\frac{\partial^2\phi }{\partial {x^{\prime}}^2} -(\frac{\omega_0^2 }{2}{x^{\prime}}^2 +a_0 {x^{\prime}})\phi =E\phi
\end{equation}
The solution of this equation is given in terms of the parabolic cylinder functions, $D_n(x)$. Transforming backwards yields the two linearly independent solutions (\ref{ufegdsdaw2b}).

\newpage

\begin{figure}[t]
\includegraphics[width=17cm]{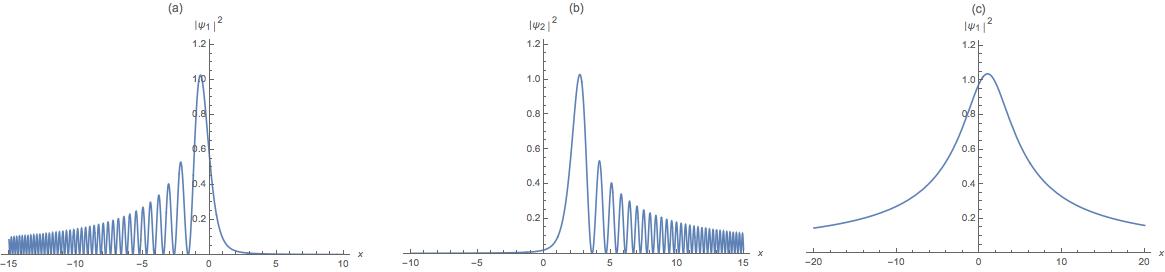}
\caption{The intensities for the two linearly independent solutions (\ref{ufegdsdaw2b}) when $E=0$ (a,b) and when $E=5$ (c) ($|\psi_1(x)|^2\approx |\psi_2(x)|^2$ for large $E$). As can be seen, $\psi_1$ and $\psi_2$ don't converge in the negative and positive branches, respectively, while a combination of these two functions does not go to zero as $\ds{x\rightarrow\mp\infty}$.  The solutions have oscillating tails only for small values of $E$. We stress that none of the wave packet is integrable.The parameters are given by $\ds{a_0=\omega_0=L=1}$ and $x_c=0$. In all figures, the maximum intensity of the main peak is set to unity.}
\end{figure}

\begin{figure}[t]
\includegraphics[width=17cm]{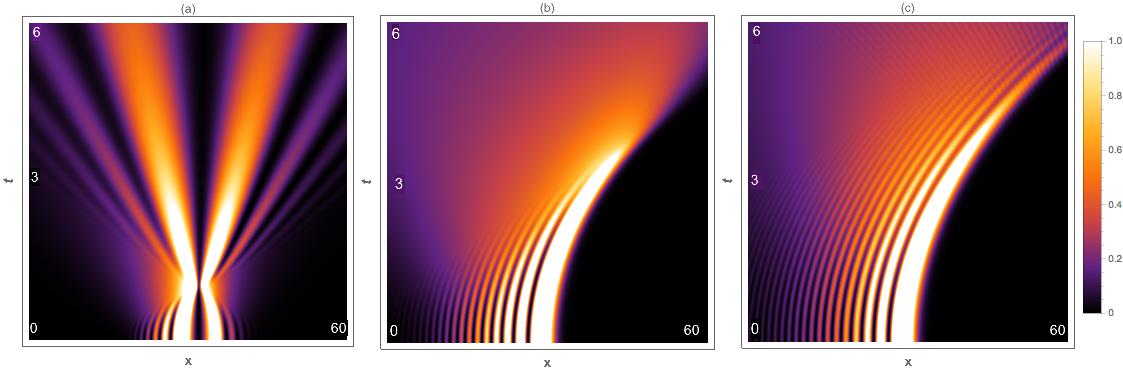}
\caption{The density plots $\ds{|\Psi(x,t)|^2}$ as a function of $x$ and $t$ are shown in the figures (a)-(c) for free particle, $\Omega^2=0$. We start with the initial wave packet $\ds{\Psi_1-\Psi_2}$ with $\ds{\epsilon=\frac{1}{20}}$ in (a) and $\ds{\Psi_2}$ with $\ds{\epsilon=\frac{1}{100}}$ in (b). In Fig. (c), the density plot can be seen for the truncated Airy wave $\ds{e^{0.1x} Ai(2^{1/3}x)}$ for comparision}
\end{figure}

\begin{figure}[t]
\includegraphics[width=17cm]{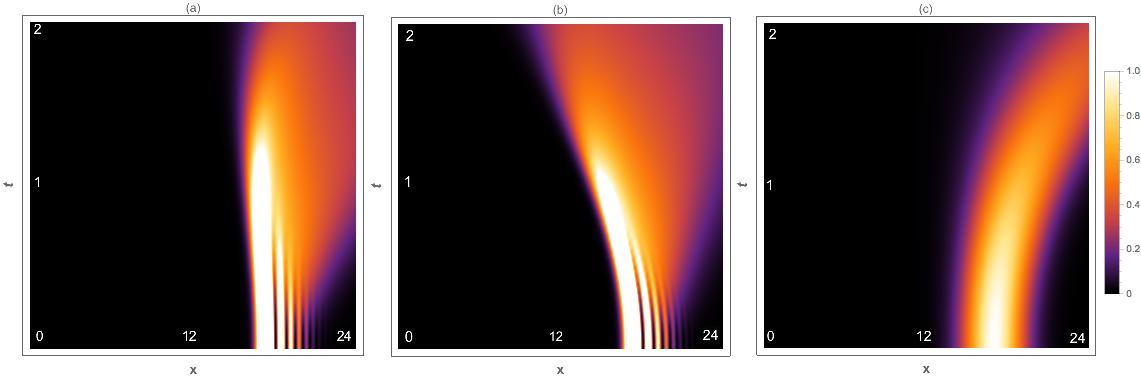}
\caption{ The densities $\ds{|\psi(x,t)|^2}$ for the accelerating and Gaussian initial wave packets in the presence of the inverted harmonic potential with the center at $x=12$. The displaced initial wave packets are given by $\ds{e^{\frac{-(x-17)^2}{10}}\psi_2(x-17)}$ with $a_0=-5,\omega_0=0.6$ and $\ds{\omega^2=0.7}$ (a) $\ds{\omega^2=0.1}$ (b) . For comparison, we plot the propagation of the Gaussian beam $\ds{e^{\frac{-(x-17)^2}{10}}}$ for $\ds{\omega^2=0.7}$ (c). The Gaussian wave packet moves towards the right hand side since the inverted potential is maximum at $x=12$ and decreases quadratically with the distance away from the center. On the other hand, the main peak of the accelerating wave packet can move towards the point where the potential is maximum. This is possible if the self-acceleration parameter $a_0$ is chosen to be negative and big enough such that self-acceleration overcomes the acceleration due to the inverted potential. The main peak of the wave packet in Fig. (a) is not moving (until $t=1$) as if there is no external potential. For a smaller value of $\omega^2$, the main peak moves towards the center of the inverted potential.  }
\end{figure}

\end{document}